\documentclass[a4paper,12pt]{article}
\pdfoutput=1
\usepackage{epsfig}
\usepackage{amssymb}
\usepackage{amsfonts}
\usepackage{amsmath}
\usepackage{euscript}
\usepackage{verbatim}
\usepackage{latexsym}
\usepackage{graphicx}
\usepackage{caption}
\usepackage{float}
\usepackage{subcaption}
\usepackage{wrapfig}
	\usepackage[T1]{fontenc}
	\usepackage{tikz}
	\usetikzlibrary{decorations.pathmorphing}
\usepackage{tikz}
\usetikzlibrary{shapes.geometric, arrows,patterns,snakes}
\tikzstyle{ellip} = [ellipse, minimum width=3cm, minimum height=1cm,text centered, draw=black]

\newskip\humongous \humongous=0pt plus 1000pt minus 1000pt

\newif\ifdtup

\catcode`\@=11

\@addtoreset{equation}{section}

\def\@normalsize{\@setsize\normalsize{15pt}\xiipt\@xiipt
\abovedisplayskip 14pt plus3pt minus3pt%
\belowdisplayskip \abovedisplayskip
\abovedisplayshortskip \z@ plus3pt%
\belowdisplayshortskip 7pt plus3.5pt minus0pt}

\def\small{\@setsize\small{13.6pt}\xipt\@xipt
\abovedisplayskip 13pt plus3pt minus3pt%
\belowdisplayskip \abovedisplayskip
\abovedisplayshortskip \z@ plus3pt%
\belowdisplayshortskip 7pt plus3.5pt minus0pt
\def\@listi{\parsep 4.5pt plus 2pt minus 1pt
     \itemsep \parsep
     \topsep 9pt plus 3pt minus 3pt}}

\relax

\catcode`@=12

\catcode`\@=11

\def\section{\@startsection{section}{1}{\z@}{3.5ex plus 1ex minus
   .2ex}{2.3ex plus .2ex}{\large\bf}}

\def\SymBoxes#1#2#3#4{\newdimen\un@t \un@t#3%
\raisebox{#1}{\rule{#2\un@t}{#4}\hskip-#2\un@t
\@tempdimb\un@t \advance\@tempdimb by-#4\@tempcntb#2\relax%
\@whilenum{\@tempcntb>0}\do{
\rule{#4}{\un@t}\hskip\@tempdimb \advance\@tempcntb by\m@ne}%
\hskip-#2\un@t \rule[\un@t]{#2\un@t}{#4}%
\rule[\un@t]{#4}{#4}\hskip-#4
\rule{#4}{\un@t}}\hskip-#4}                

\begin{document}

\newcommand{\beq}{\begin{equation}}
\newcommand{\eeq}{\end{equation}}
\newcommand{\bea}{\begin{eqnarray}}
\newcommand{\eea}{\end{eqnarray}}
\newcommand{\beas}{\begin{eqnarray*}}
\newcommand{\eeas}{\end{eqnarray*}}
\newcommand{\defi}{\stackrel{\rm def}{=}}
\newcommand{\non}{\nonumber}
\newcommand{\bquo}{\begin{quote}}
\newcommand{\enqu}{\end{quote}}
\renewcommand{\(}{\begin{equation}}
\renewcommand{\)}{\end{equation}}
\def \eqn#1#2{\begin{equation}#2\label{#1}\end{equation}}
\def\IZ{{\mathbb Z}}
\def\IR{{\mathbb R}}
\def\IC{{\mathbb C}}
\def\IQ{{\mathbb Q}}
\def\de{\partial}
\def\Tr{ \hbox{\rm Tr}}
\def\H{ \hbox{\rm H}}
\def\HE{ \hbox{$\rm H^{even}$}}
\def\HO{ \hbox{$\rm H^{odd}$}}
\def\K{ \hbox{\rm K}}
\def\Im{ \hbox{\rm Im}}
\def\Ker{ \hbox{\rm Ker}}
\def\const{\hbox {\rm const.}}
\def\o{\over}
\def\im{\hbox{\rm Im}}
\def\re{\hbox{\rm Re}}
\def\bra{\langle}\def\ket{\rangle}
\def\Arg{\hbox {\rm Arg}}
\def\Re{\hbox {\rm Re}}
\def\Im{\hbox {\rm Im}}
\def\exo{\hbox {\rm exp}}
\def\diag{\hbox{\rm diag}}
\def\longvert{{\rule[-2mm]{0.1mm}{7mm}}\,}
\def\a{\alpha}
\def\dag{{}^{\dagger}}
\def\tq{{\widetilde q}}
\def\p{{}^{\prime}}
\def\W{W}
\def\N{{\cal N}}
\def\hsp{,\hspace{.7cm}}

\def\br{\nonumber\\}
\def\IZ{{\mathbb Z}}
\def\IR{{\mathbb R}}
\def\IC{{\mathbb C}}
\def\IQ{{\mathbb Q}}
\def\IP{{\mathbb P}}
\def \eqn#1#2{\begin{equation}#2\label{#1}\end{equation}}

\newcommand{\sgm}[1]{\sigma_{#1}}
\newcommand{\idd}{\mathbf{1}}

\newcommand{\C}{\ensuremath{\mathbb C}}
\newcommand{\Z}{\ensuremath{\mathbb Z}}
\newcommand{\R}{\ensuremath{\mathbb R}}
\newcommand{\rp}{\ensuremath{\mathbb {RP}}}
\newcommand{\cp}{\ensuremath{\mathbb {CP}}}
\newcommand{\vac}{\ensuremath{|0\rangle}}
\newcommand{\vact}{\ensuremath{|00\rangle}                    }
\newcommand{\oc}{\ensuremath{\overline{c}}}
\begin{titlepage}
\begin{flushright}
CHEP XXXXX
\end{flushright}
\bigskip
\def\thefootnote{\fnsymbol{footnote}}

\begin{center}
{\Large
{\bf Random Matrices and Holographic Tensor Models
}
}
\end{center}

\bigskip
\begin{center}
{\large  Chethan KRISHNAN$^a$\footnote{\texttt{chethan.krishnan@gmail.com}}, K. V. Pavan KUMAR$^a$\footnote{\texttt{kumar.pavan56@gmail.com}}, \vspace{0.15in} \\ and \ Sambuddha SANYAL$^b$\footnote{\texttt{sambuddha.sanyal@icts.res.in}} }
\vspace{0.1in}

\end{center}

\renewcommand{\thefootnote}{\arabic{footnote}}

\begin{center}
$^a$ {Center for High Energy Physics,\\
Indian Institute of Science, Bangalore 560012, India}

$^b$ {International Center for Theoretical Sciences,\\
Tata Institute of Fundamental Research, Bangalore 560089, India}\\

\end{center}

\noindent
\begin{center} {\bf Abstract} \end{center}

We further explore the connection between holographic $O(n)$ tensor models and random matrices. First, we consider the simplest non-trivial uncolored tensor model and show that the results for the density of states, level spacing and spectral form factor are qualitatively identical to the colored case studied in arXiv:1612.06330. We also explain an overall 16-fold degeneracy by identifying various symmetries, some of which were unavailable in SYK and the colored models. Secondly, and perhaps more interestingly, we systematically identify the Spectral Mirror Symmetry and the Time-Reversal Symmetry of both the colored and uncolored models for all values of $n$, and use them to identify the Andreev ensembles that control their random matrix behavior. We find that the ensembles that arise  exhibit a refined version of Bott periodicity in $n$.


\vspace{1.6 cm}
\vfill

\end{titlepage}

\setcounter{footnote}{0}


\section{Introduction}

The Sachdev-Ye-Kitaev (SYK) model \cite{SYK, Polchinski, maldacena, old, coredump} has recently emerged as a a candidate for a 0+1 dimensional holographic model for black holes\cite{qftbh}. SYK involves a disorder averaging, and for some purposes (especially for investigations on black hole unitarity), it might be useful to have theories that have the same large-$N$ structure as SYK, but do not need a disorder average. Tensor models which exhibit this feature have been constructed \cite{witten, guraudump, klebanov, tanasa}, and these Holographic Tensor Models are the topic of this paper. 

A specific example of a colored tensor model was recently explicitly diagonalized \cite{bala} and used to demonstrate features of random matrices and chaos (as expected of black holes). It exhibited striking parallels with SYK \cite{cotler}, but also some differences. One difference was that the spectrum showed huge accidental degeneracies\footnote{By accidental degeneracy, we mean degeneracies that depended on the energy level. The overall $m$-fold degeneracies that could be directly understood in terms of the symmetries of the Hamiltonian are a simpler issue.} which did not exist in (a single sample of) SYK. The spectrum also was not as rigid as it was in SYK, and had gaps in it. The final qualitative difference was that the Hamiltonian had a discrete symmetry\footnote{More precisely, a {\em pseudo-}symmetry: there exists a unitary operator that anti-commutes with the Hamiltonian.} because of which the random matrix ensemble that controlled its chaos-like behavior was unlikely to be one of the Wigner-Dyson (GOE/GUE/GSE) ensembles. Instead it belonged to the so-called BDI class in the Andreev-Altland-Zirnbauer classification of random matrix ensembles. 

In this paper we would like to understand the robustness/genericity of the above observations, which were limited to a single case in \cite{bala}. It should be clear that the explicit diagonalization approach is not going to be of further mileage in colored models with larger $N$, because we were already at the limits of computability in \cite{bala} as was already emphasized there. So we proceed simultaneously in two ways. 

Firstly, we consider the simplest {\em un-}colored models (the cases $q=4, d=2$ and $q=4, d=3$ in \cite{klebanov}) and {\em numerically} diagonalize them and explore their random-matrix like features repeating the approach of \cite{bala}. The $q=4, d=2$ case turns out to be too small to exhibit any chaos/random matrix behavior\footnote{This is not surprising, this system is comparable to the $N=8$ Majorana SYK case.}. But we find that {\em all} the features that we saw in \cite{bala} have incarnations in the $q=4, d=3$ case as well. In particular, the density of states has a similar form, there is level repulsion, and the spectral form factor exhibits a dip-ramp-plateau structure (after a running time average). Even more strikingly, the large accidental degeneracies, the gaps in the spectrum and the lack of rigidity are still there\footnote{We emphasize however that these results are for relatively small $N$, so it is unclear if they persist as $N \rightarrow \infty$.}, and so is the spectral mirror symmetry.

Secondly, since numerical diagonalization has no traction at generic $n$, we {\em analytically} identify the discrete (pseudo-)symmetries of both the colored and uncolored tensor models for arbitrary $n$. We find that they are again distinct from the Wigner-Dyson ensembles: in fact more classes beyond the BDI class of \cite{bala} show up, and we identify them as a function of $n$. We note the presence of a Bott-like periodicity in the symmetry class as a function of $n$, which has some parallels to the one found in \cite{xu} for SYK.

\section{SYK vs (Un)colored Holographic Tensor Models}

In this section, we will compare and contrast the three models- SYK model, (Gurau-Witten) colored tensor model and the uncolored tensor model. Note that we discuss only the salient features of these models and for more details, the reader is referred to the original papers.

Sachdev-Ye-Kitaev model is a 0+1-dimensional model of fermions in which the interaction terms involve an even number of fermions, typically four. The coupling constants are picked from a Gaussian distribution. While computing the correlators in the SYK model, we need to take an average over these random couplings. This disorder averaging\footnote{Disorder averaging cannot be avoided if one wants the large-$N$ solvability of SYK.} is avoided in the tensor models, while retaining the large-$N$ behavior of SYK.

To address the above issue, Witten proposed a model \cite{witten} based on the work of Gurau on colored tensor models \cite{guraudump}. Gurau-Witten model is dependent on two independent integers-- $d$ and $n$. The basic building blocks of the model are the fermionic tensors of the form $\psi ^{{i_1}{i_2}\ldots {i_d}}_A$. The index $A$ corresponds to color and runs from $0$ to $d$ whereas the indices (${i_1}\ldots {i_d}$) take the values from $1$ to $n$. Hence, the degrees of freedom i.e., the number of independent fermionic fields in this model is $N=(d+1)n^d$.

The interaction term in Gurau-Witten model consists of $q=d+1$ fermions and the contractions of various tensorial indices are done following the prescription given in \cite{witten}. For every distinct pair ($A,B$) belonging to ($0,\ldots d$), we assign a symmetry group $O(n)$ and hence the overall symmetry group\footnote{To be precise, there is a quotient with respect to a discrete group. But this does not affect our discussions and hence we ignore it in the paper.} ($G$) of the theory is a direct product of all the $O(n)$'s leading us to
\begin{align}
G\sim [O(n)]^{d(d+1)/2}
\end{align}
The interaction term one writes, should be an invariant\footnote{More precisely, we demand that for any given $O(n)$, there are exactly two of the fermionic tensors that transform as vectors and rest of them would transform as scalars.} under $G$. To put it in an operationally convenient way, for any two distinct fermions $\psi _A$ and $\psi _B$, there should be exactly one tensorial index that is contracted. Further, since we take  the interaction term containing $q=d+1$ fermions, it follows that each of the fermions has contractions with every other fermion in the interaction term. 

One of the reasons to consider the above model is that the  large $N$ perturbation theory of Gurau-Witten model has a similar behaviour as that of SYK model. In particular, in both cases, the melonic diagrams dominate in the large $N$ limit. Further, both the models are maximally chaotic i.e, they saturate the Maldacena-Shenker-Stanford bound \cite{bound}. Also, one of the major advantages of Gurau-Witten model is that gauging the symmetry group $G$ would allow us to construct gauge singlets, which have a well-defined dual interpretation in the bulk. Gurau-Witten model, therefore, combines the properties of large $N$ solvability and maximally chaotic nature of SYK model with a possibility of defining the bulk dual using the currently available technology. Hence, this model could serve as a prototype to study quantum black holes.

 It was suggested by Klebanov and Tarnopolsky (based on \cite{tanasa}) that one could construct a tensor model \cite{klebanov} with a smaller symmetry group that still enjoys the salient features of Gurau-Witten model. This new model is an uncolored tensor model in the sense that we do not distinguish among the fermions and build the entire model using a single fermionic tensor $\psi ^{{i_1}{i_2}\ldots {i_d}}$. We will still distinguish the tensorial indices though and as in the Gurau-Witten model, each $i_m$ can take values from $1$ to $n$. So, the degrees of freedom in this model is $N=n^d$.

The theory has $G_u\sim [O(n)]^d$ symmetry coming from the transformation properties of ${i_m}$'s. The fermionic tensors $\psi ^{{i_1}{i_2}\ldots {i_d}}$ transform under $G_u$ as:
\begin{align}
\psi ^{{i_1}{i_2}\ldots {i_d}}\rightarrow G_u^{{i_1}{j_1}}~G_u^{{i_2}{j_2}}\ldots G_u^{{i_d}{j_d}} ~\psi ^{{j_1}{j_2}\ldots {j_d}}
\end{align}
The interaction term for the uncolored model needs to be an invariant under $G_u$. Since the symmetry group is $[O(n)]^d$, we need to have at least $d+1$ fermions in the interaction term to form an invariant. Klebanov and Tarnopolosky have considered the interaction term involving $d+1$ fermions in their paper and we will follow that. As there are only $d+1$ fermions in the interaction term, there exists a contraction between any two distinct fermions, analogous to Gurau-Witten model.   

The Gurau-Witten model was diagonalized explicitly in \cite{bala}, in the next section we will do the same thing for the uncolored model. 

\section{$n=3$; $d=3$ Uncolored Tensor Model}

The simplest interacting case of the uncolored model arises when we take $n=2$ and $d=3$, which leads to $N=8$. As explained in \cite{bala}, we can assign a gamma matrix of $SO(N)$ to each of the $N$ fermions $\psi ^{{a_1}{a_2}...}$. Hence, the Hamiltonian in this case is a 16-dimensional matrix. Even though the number of eigenvalues are not high enough to demonstrate signs of quantum chaos, the eigenvalue spectrum has some interesting properties. The spectrum has a degeneracy of 14 at the mid-level energy. The other two levels are placed equidistantly on either side of the mid-point.

Now, we consider the next non-trivial case of $n=3$ and $d=3$ which gives rise to $N=27$ \footnote{The next case is too big for our computers, but fortunately we find that this case already exhibits features of chaos.}. Even though $N=27$, we choose\footnote{We work with $SO(28)$ gamma matrices even though we could work with $SO(27)$, because that is conventional in the condensed matter literature. The motivation for this seems to be that one can think of the Hilbert space as being generated {\em perturbatively} by (complex) creation and annihilation operators which are made from pairs of Majoranas. We observe that working with $SO(27)$ instead of $SO(28)$ leads to halving of degeneracy at all levels. It is worth noting that an odd number of Majorana fermions has a gravitational anomaly, even though this is not particularly important since we never couple our system to gravity. } to work with gamma matrices of $SO(28)$ instead of $SO(27)$. Hence the Hamiltonian is a 16384-dimensional matrix. Following the rules of contractions given in the last section, we can write the Hamiltonian as\footnote{In the rest of the paper, we set the parameter $J=1$.}  

\begin{align}
\label{hamiltonian}
H&=\frac{J}{\sqrt{8}}\sum _{{a_1},{a_2}}\sum _{{b_1},{b_2}} \sum _{{c_1},{c_2}} \psi ^{{a_1}{b_1}{c_1}}\psi ^{{a_1}{b_2}{c_2}}\psi ^{{a_2}{b_1}{c_2}}\psi ^{{a_2}{b_2}{c_1}}
\end{align}
where each of the indices $(a_1,a_2,b_1,b_2,c_1,c_2)$ runs from $1$ to $3$. Also, we assign the gamma matrices of $SO(28)$ to the fermions as follows:
\begin{align}
\psi ^{ijk}&=\gamma _{p} \\
p&=9(i-1)+3(j-1)+k
\end{align} 
Note that only first 27 gamma matrices of $SO(28)$ are present in the Hamiltonian. Also, we will not write the explicit form of Hamiltonian because it contains 729 terms. In fact, for a generic choice of $n$ and $d$, the Hamiltonian of both the colored and uncolored models consists of $n^{d(d+1)/2}$ terms.

\begin{figure}
\centering
\includegraphics[scale=0.5]{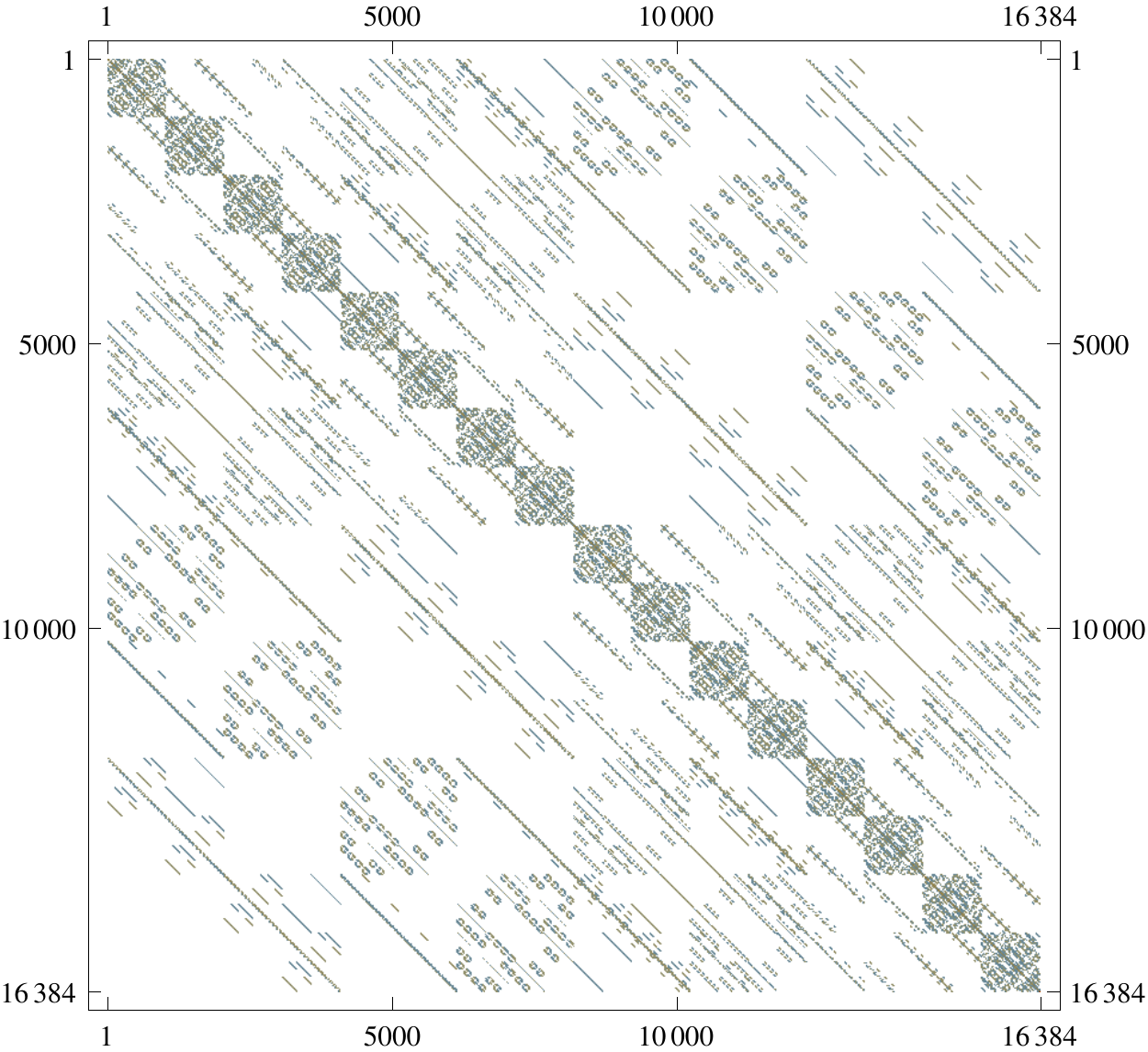}
\caption{MatrixPlot of the Hamiltonian \eqref{hamiltonian}}
\label{hamplot}
\end{figure}

Once we have the relation between gamma matrices\footnote{We refer the reader to the appendix regarding the details of our choice of  representation for gamma matrices.} and fermions, we can proceed to find the eigenvalues of the Hamiltonian\footnote{Despite  some of the gamma matrices having imaginary entries and the Hamiltonian containing 729 terms, the Hamiltonian \eqref{hamiltonian} is both real and symmetric.}. Before actually finding the eigenvalues, it is a good idea to get an understanding on how the matrix sparseness structure of Hamiltonian looks like. We plot this in figure \ref{hamplot}. Comparing figure \ref{hamplot} with the matrix plot of $n=2$; $d=3$ Hamiltonian of Gurau-Witten model (see \cite{bala}), we see that the plot for uncolored model is denser \footnote{See appendix \ref{sparsity} for details.}. This can be attributed to the higher number of terms in the Hamiltonian in the present case.

Our goal is to perform the analysis (similar to \cite{bala}) of the eigenvalue spectrum of the above Hamiltonian \eqref{hamiltonian} and to verify whether the uncolored model showcases the features of quantum chaos that are present in Gurau-Witten model. 

\subsection{Eigenvalue Spectrum}

\begin{figure}
\centering
\includegraphics[scale=1.2]{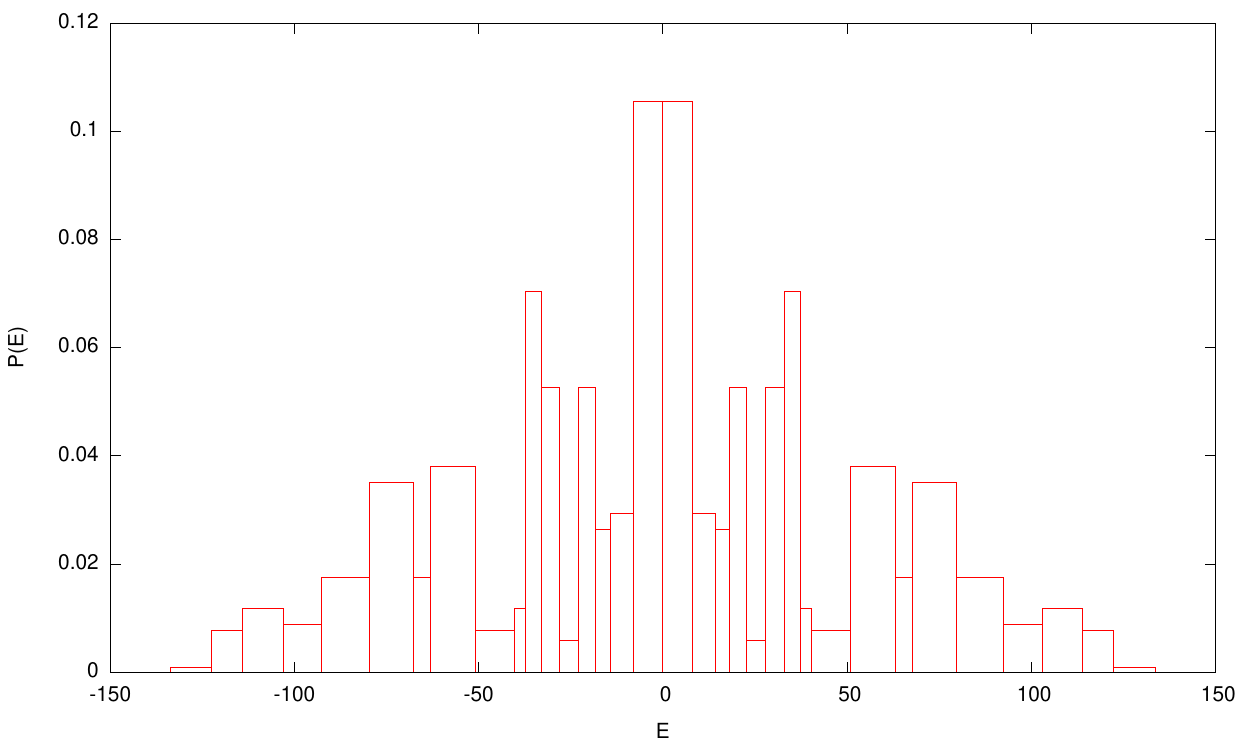}
\caption{The density of states vs the energy levels. The spectrum has \textit{spectral mirror symmetry} i.e., the spectrum is symmetric around $E=0$.}
\label{DOS}
\end{figure}

To start with, we give the details of the eigenvalues of the Hamiltonian \eqref{hamiltonian}. Upon diagonalizing the Hamiltonian numerically, we find that the spectrum has \textit{spectral mirror symmetry} i.e., the spectrum is symmetric about the mid-level energy. To make the entire spectrum symmetric about $E=0$, we shift the entire spectrum by a constant. To get an intuition on how the spectrum looks like, the density of states with respect to the energy levels is plotted in figure \ref{DOS}.

\begin{figure}
\centering
\includegraphics[scale=0.9]{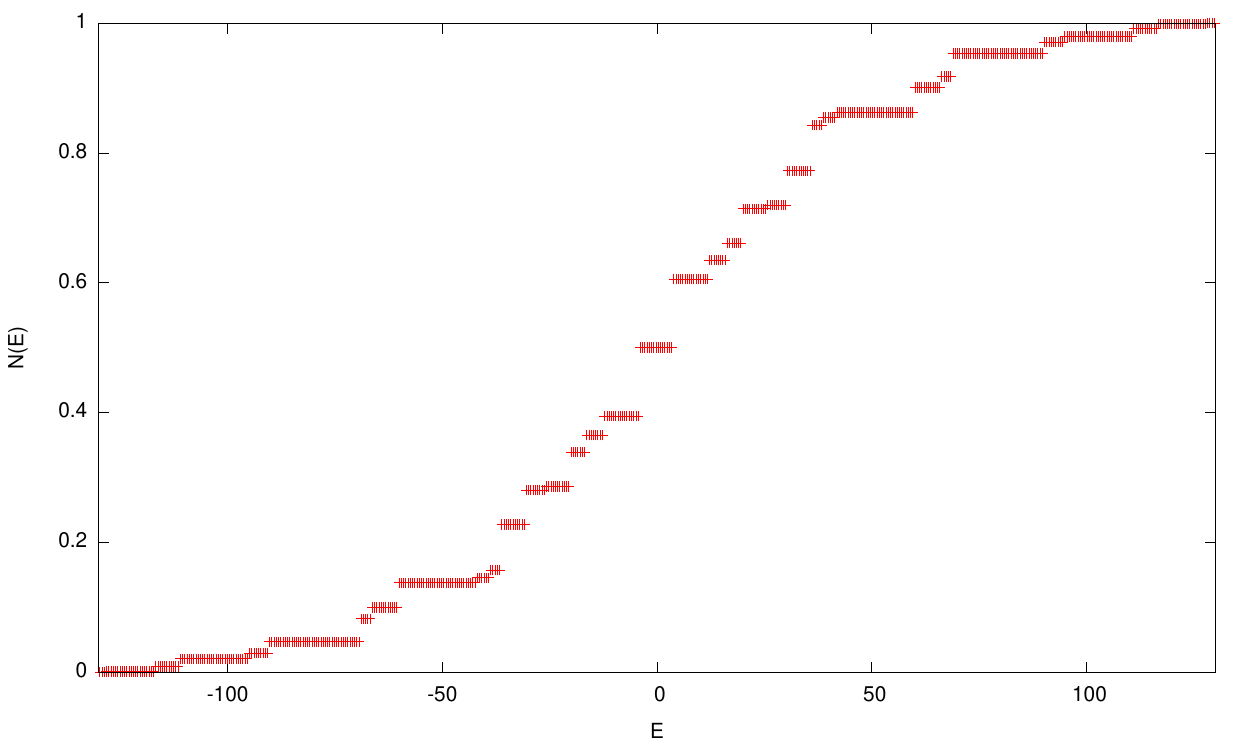}
\caption{The integrated density of states vs the energy levels}
\label{IDOS}
\end{figure}

Further, we also plot the integrated density of states as a function of energy in figure \ref{IDOS}. Upon a quick look at the figure, one can notice that the \textit{accidental} degeneracies present in \cite{bala} make an appearance in the uncolored model too. We do not completely understand the origin or the significance of these accidental degeneracies. But, we note that there is a 16-fold degeneracy associated to all the energy levels of the spectrum and we can explain this degeneracy by understanding the symmetries of the Hamiltonian. We will do this in section \ref{degeneracies}.

We note that after removing the degeneracies, there are only 34 eigenvalues left. This is a smaller number of independent eigenvalues than in \cite{bala}, but we will find that the chaos/random matrix features we found there are fairly robustly reproduced here as well.

\subsection{Spectral Form Factor}

Spectral form factor (SFF)  encodes information regarding the structure of eigenvalues of a system. In \cite{cotler}, Cotler et al  compared the SFF computed using a random matrix chosen from one\footnote{Gaussian Unitary Ensemble (GUE), to be precise.} of the Dyson ensembles with that of SYK model and concluded that the running time average plot of SFF in both cases have a similar behaviour. In particular, both the plots have a dip, a ramp and a plateau. The late time plateau is related to level repulsion. It was shown in \cite{bala} that the SFF plot of simplest non-trivial case of Gurau-Witten model admits the dip-ramp-plateau behaviour (qualitatively) and hence suggests random matrix/chaos behaviour. We will show that this is also the case for the uncolored models. The physics of the dip-ramp-plateau structure is discussed in detail in \cite{cotler}, see also the introduction of \cite{bala} for a short overview.

Spectral form factor is defined to be 
\begin{align*}
F_{\beta }(t)&=\left|\frac{Z(\beta ,t)}{Z(\beta )}\right|^2
\end{align*} 
where $Z$ is the partition function and is defined to be
\begin{align*}
Z(\beta ,t)&=\text{Tr}\left(e^{-(\beta +i t)H}\right)
\end{align*}

\begin{figure}
\centering
\includegraphics[scale=1]{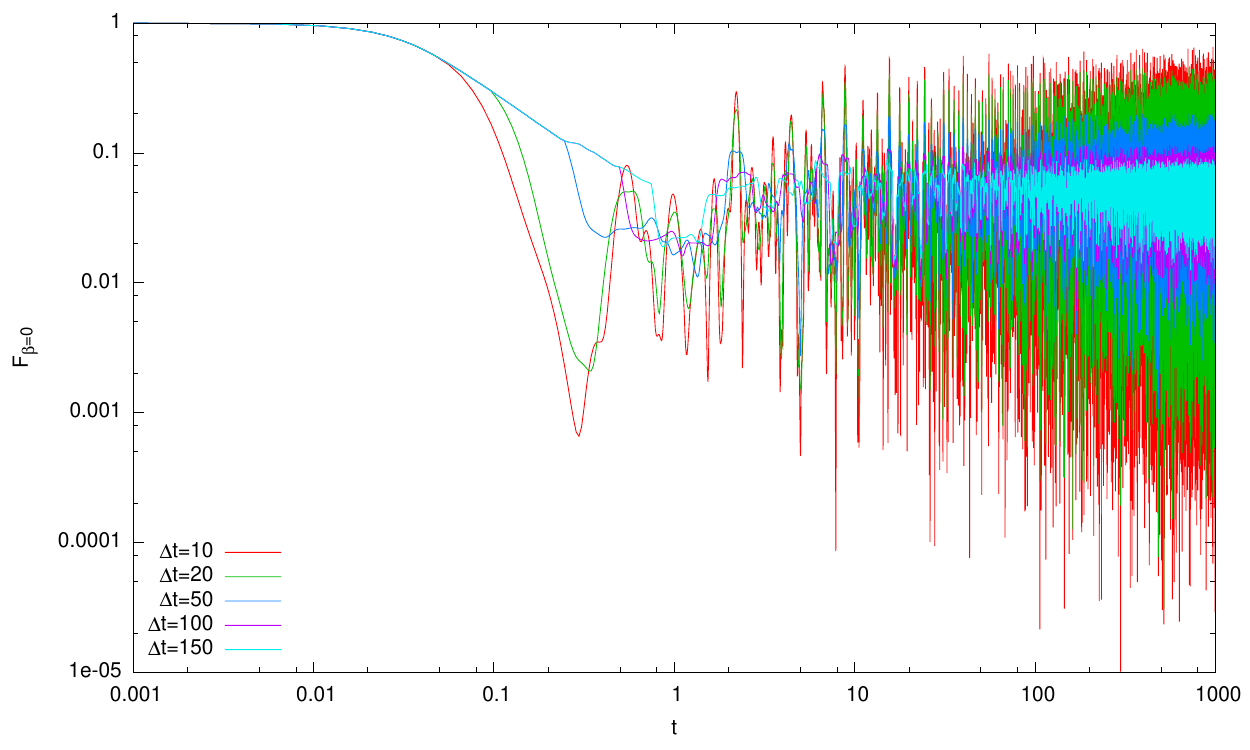}
\caption{SFF for $\beta =0$ for various window sizes ($\Delta t$). Similar to SYK and Gurau-Witten models, there is a dip-ramp-plateau structure present here as well, suggesting a random matrix/chaotic behaviour }
\label{SFF_0}
\end{figure}

\begin{figure}
\centering
\includegraphics[scale=1]{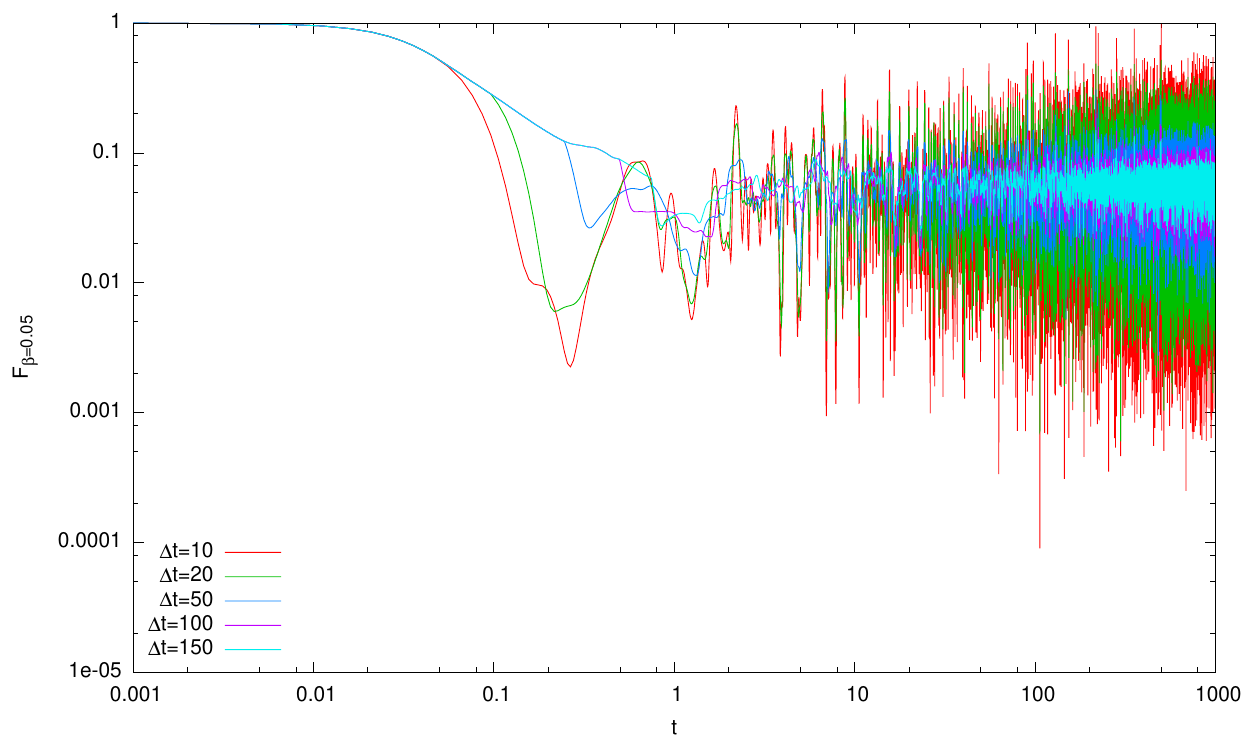}
\caption{SFF for $\beta =0.05$ for various window sizes ($\Delta t$)}
\label{SFF_p05}
\end{figure}

For the $n=3$; $d=3$ uncolored model, we have computed SFF for various values of $\beta $. Then, for a couple of $\beta $'s, we have plotted the SFF after a sliding average with various window sizes ($\Delta t$) that are listed in the plots \ref{SFF_0} and \ref{SFF_p05}. We notice that the SFF plots in our case have similar qualitative features as that of SYK and Gurau-Witten model. The dip-ramp-plateau structure is distinct when the inverse temperature $\beta \rightarrow 0$ but gets more messy as we increase $\beta $. We expect that this is due to the smallness of the $N=27$ case and the associated low number (34) of eigenvalues present once we have removed the degeneracies. In any event, it is evident that the structure that was seen in \cite{bala} is present here as well.

\subsection{Level Repulsion}

One of the major signatures of quantum chaos is the presence of level repulsion in the level spacing plot. In classically integrable systems, the distribution of energy level spacings $P(s)$ mimics a Poisson distribution (See \cite{berry2}, for example) and hence the peak of the distribution occurs as $s\rightarrow 0$. Equivalently, the spectrum of an integrable system predominantly includes levels that are closely separated. But, for a chaotic system, one observes a turn around near $s\rightarrow 0$ (see figure 8 in \cite{bala}, for example) and the energy levels tend to separate away from each other. This phenomenon is called level repulsion. 

\begin{figure}
\centering
\includegraphics[scale=1]{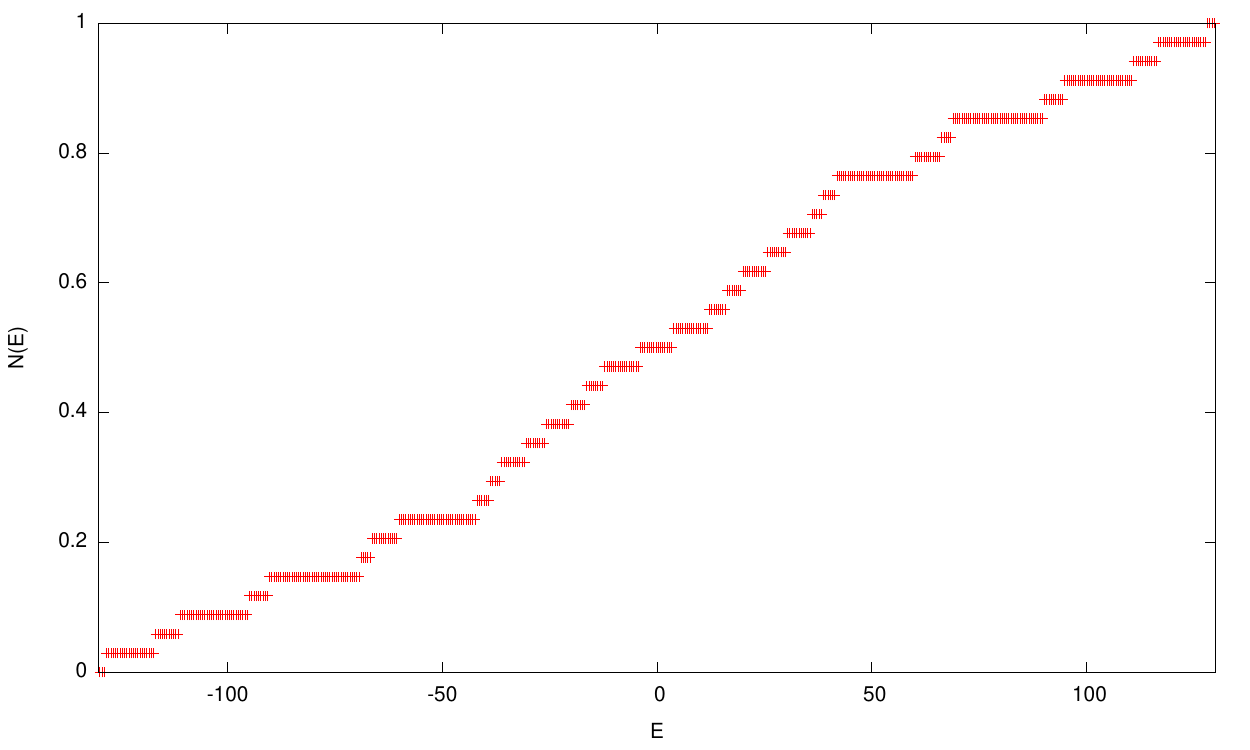}
\caption{Integrated Density of States after removing the degeneracies}
\label{IDOS_nondeg}
\end{figure}

\begin{figure}
\centering
\includegraphics[scale=1]{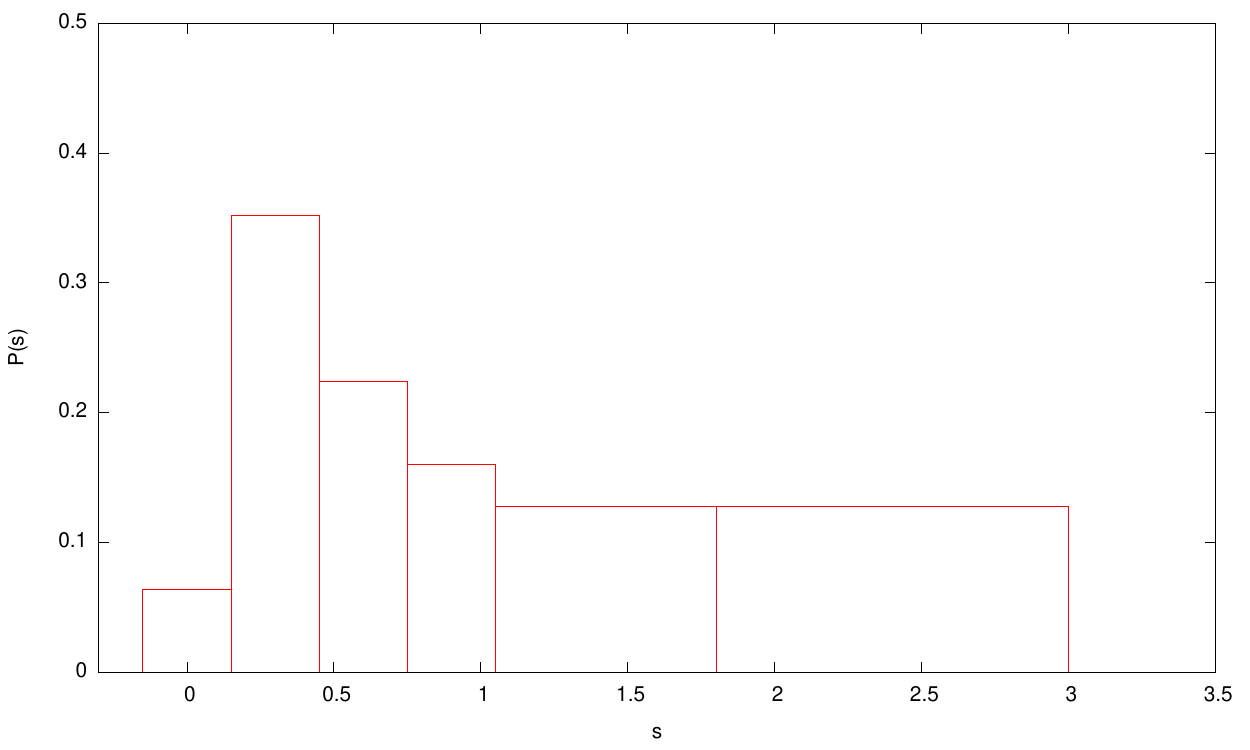}
\caption{Level spacing distribution for the unfolded spectrum. The vertical axis $P(s)$ has been scaled for convenience and this does not affect any of our discussions as we are mainly interested in turn around near $s\rightarrow 0$. This turn around suggests level repulsion.}
\label{level spacing}
\end{figure}

To plot the level spacing distribution, we first remove the degeneracies among the eigenvalues and retain only the distinct energy levels. The non-degenerate spectrum is shown in figure \ref{IDOS_nondeg}. 
Then, the spectrum is \textit{unfolded}\footnote{ The unfolded spectrum is usually given by a function of the original spectrum such that the mean level separation between various levels is unity. The unfolding process helps in smoothing out the inhomogeneities of the spectrum. See chapter 4 of \cite{haake} for more details.} following \cite{garcia} and the level spacing distribution is plotted in figure \ref{level spacing} using the unfolded data. Even though we have only 34 eigenvalues after removing the degeneracies, the level repulsion is manifest in figure \ref{level spacing}. This behaviour is similar to that of \cite{bala}. 

\section{Discrete Symmetries of the Hamiltonian}
\label{degeneracies}

As we have seen in the previous section, the eigenvalues of the Hamiltonian possess the following features:
\begin{itemize}
\item The eigenvalue spectrum has \textit{spectral mirror symmetry} i.e., the spectrum is symmetric around the mid-level energy
\item All the energy levels are at least 16-fold degenerate i.e., this 16-fold degeneracy can be understood by studying the symmetries of the Hamiltonian
\end{itemize}
This section is devoted to explaining the above features of the spectrum. In particular, we start by identifying an operator $S$ that anticommutes with the Hamiltonian to explain the spectral mirror symmetry. Then we move on to finding various symmetries of the Hamiltonian that are responsible for the 16-fold degeneracy.

\subsection{Spectral Mirror Symmetry}

To begin with, we remind ourselves that the Hamiltonian we are working with, is of the form 
\begin{align}
\label{hamiltonian-repeat}
H&=\sum _{a_1,a_2}\sum _{b_1,b_2}\sum _{c_1,c_2}\psi ^{a_1b_1c_1}\psi ^{a_1b_2c_2}\psi ^{a_2b_1c_2}\psi ^{a_2b_2c_1}
\end{align}
with each of $\{a_1,a_2,b_1,b_2,c_1,c_2\}$ taking values from 1 to $n$ (=3 here). Each fermionic field is assigned with a gamma matrix as follows:
\begin{align}
\label{gamma assignments}
\psi ^{ijk}&=\gamma ^{9i+3j+k-12} 
\end{align}
Once this assignment is done, it is straightforward to notice that, if any two gammas are the same, then the remaining two also should be equal to each other due to the contraction structure of the Hamiltonian. This observation implies that whenever one of the following conditions is met:
\begin{itemize}
\item $a_1=a_2$; $b_1=b_2$; $c_1\neq c_2$
\item $b_1=b_2$; $c_1=c_2$; $a_1\neq a_2$
\item $c_1=c_2$; $a_1=a_2$; $b_1\neq b_2$
\item $a_1=a_2$; $b_1=b_2$; $c_1= c_2$
\end{itemize}
then the corresponding term in the Hamiltonian is an identity matrix.  For a generic $n$, the Hamiltonian consists of $n^3(n-1)^2(n+2)$ such identity terms. Of the above conditions, the third condition gives an identity matrix with a negative sign and the rest of the conditions correspond to positive identity matrices. Thus, in the Hamiltonian, only $n^4$ of the total identity terms contribute and hence $n^4$ is the datum upon which the eigenvalues exhibit a mirror symmetry. Now, we separate out these identity terms in the Hamiltonian and define $H'$ as follows:
\begin{align}
H'&=H-n^4 \textbf{I}
\end{align}  
As we explained earlier, if any two gammas are the same, then the other two are also equal to each other and thus leading to an identity matrix. But, by definition, $H'$ is devoid of any such identity matrices. Hence,  all the four gamma matrices present in each of the terms of $H'$ are different from each other.

To explain the mirror symmetry, we need to find an operator $S$ such that it satisfies:
\begin{align}
SH'S^{-1}&=-H'
\end{align}
It might seem a formidable task to find such an operator as there are $\frac{n^3}{4}(n^3-3n+2)=135$ different terms in $H'$. To simplify things\footnote{We note that the operator $S$ we construct here is not unique.}, we demand that the effect of $S$ is such that it exchanges the last two gamma matrices while keeping the first two fixed at their positions. This demand is satisfied, if $S$ obeys the following condition\footnote{In principle, we could choose either sign for the action of $S$ but in this subsection, we proceed with the negative sign.}:
\begin{align}
\label{condition on S}
S(\gamma ^{9i+3j+k-12})S^{-1}&=-\gamma ^{9i+3k+j-12}
\end{align}
In principle, it is a (non-trivial) reassignment\footnote{ Equivalently, it can be interpreted as a change of representation of gamma matrices.} of gamma matrices to the fermionic fields. This reassignment, equivalently, can be thought of as writing the Hamiltonian as:
\begin{align}
H_{S}&=\sum _{a_1,a_2}\sum _{b_1,b_2}\sum _{c_1,c_2}\psi ^{a_1c_1b_1}\psi ^{a_1c_2b_2}\psi ^{a_2c_2b_1}\psi ^{a_2c_1b_2}
\end{align}
with the gamma matrix assignments as in \eqref{gamma assignments}. By comparing the contraction structure of this Hamiltonian with that of \eqref{hamiltonian-repeat}, we see that the $H'$ parts of the two Hamiltonians differ in sign whereas the contributions from identity terms are same in both cases. 

So, what is the operator $S$ explicitly? From the above condition \eqref{condition on S}, it is straightforward\footnote{We provide a more general discussion of `$S$' operator at the beginning of section \ref{Uncolored ensembles}. } to write down the operator $S$ as:
\begin{align}
S=\left(\frac{1}{2^{9/2}}\right)&\gamma _1(\gamma _2+\gamma _4)(\gamma _3+\gamma _7)\gamma _5(\gamma _6+\gamma _8)\gamma _9\gamma _{10}(\gamma _{11}+\gamma _{13})(\gamma _{12}+\gamma _{16})\nonumber \\
&\gamma _{14}(\gamma _{15}+\gamma _{17})\gamma _{18}\gamma _{19}(\gamma _{20}+\gamma _{22})(\gamma _{21}+\gamma _{25})\gamma _{23}(\gamma _{24}+\gamma _{26})\gamma _{27}
\end{align} 
We can readily check that $S$ is unitary and also that it squares to $-1$ i.e.,
\begin{align}
S^2&=-1
\end{align}
Also, one can verify that 
\begin{align}
S(n^4\textbf{I}+H')S^{-1}&=n^4\textbf{I}-H'
\end{align}
This condition suggests that one can start with two different assignments/representations of gamma matrices and obtain eigenvalues of the form $n^4+E_i$ for one of the assignments and $n^4-E_i$ for the other. But, we know that the assignment/ representation of gamma matrices should not affect any physical results. Hence, we can conclude that the values $E_i$ should be symmetric around zero. Equivalently, the spectrum of this Hamiltonian should possess a spectral mirror symmetry around $n^4$ and indeed this is what we find numerically.

\subsection{Degeneracies of the eigenvalue spectrum}
\label{degeneracies of uncolored models}

We will explain the degeneracies of the spectrum by identifying various discrete symmetries.  How many such operators are required? First of all, we checked numerically for small values of odd $N$ that SYK models exhibit a four fold degeneracy. This four-fold degeneracy was explained in terms of discrete symmetries in \cite{cotler, xu}. Further, since the uncolored Hamiltonian has a 16-fold degeneracy, we need to find more discrete symmetry operators unique to uncolored models that commute with the Hamiltonian. From now on, we will specialize to the $N=27$ case but the generalizations to higher $N$'s are straightforward and we will discuss them in the subsequent section. Our discussions in this sub-section are partially based on \cite{cotler, xu}.  

We start by identifying the operators that commute with the Hamiltonian. Following \cite{xu}, we define a fermion parity operator $P$ as:
\begin{align}
P&=(-i)^{14}\gamma _1\gamma _2\ldots \gamma _{28}
\end{align}
Note that $P$ anticommutes with all the fermionic fields in the theory. Since all the fermions anticommute with $P$, it is easy to see that $P$ commutes with the Hamiltonian. For the case of odd $N$, we can also define an operator $Z$ as:
\begin{align}
Z&=(-i)\gamma _1\gamma _2\ldots \gamma _{27}
\end{align}
The operator $Z$ anticommutes with $P$. Also, $Z$ commutes with all the fermions and hence it commutes with the Hamiltonian. We note that $Z$ is a symmetry that is unique to odd $N$ cases. 

Next, we define a time-reversal operator $\cal T$ which is an anti-unitary operator and can be written as a product of a unitary operator ${\cal U}_T$ and a complex conjugation operator $\cal K$, which is anti-unitary. The explicit form of $\cal T$ in the current case is:
\begin{align}
{\cal T}&=-\gamma _2\gamma _4\ldots \gamma _{28}{\cal K}
\end{align}
We choose our conventions such that $\cal K$ anticommutes with even gamma matrices and commutes with odd gamma matrices i.e.,
\begin{align}
{\cal K}\gamma _{a}{\cal K}^{-1}&=-(-1)^a \gamma _a
\end{align}
Upon using this property of $\cal K$ along with the explicit form of $\cal T$, we find that the time reversal operator commutes with all the gamma matrices and hence we obtain the following:
\begin{align}
[T,H]&=0 \\
[T,P]&=0 \\
\{T,Z\}&=0
\end{align}
The anticommutation of time reversal operator and the operator $Z$ is because of the presence of ``$i$'' in the definition of $Z$. Further, we can compute the value of ${\cal T}^2$ to be $-1$ i.e.,
\begin{align}
{\cal T}^2&=-1
\end{align}    
As both $S$ and ${\cal T}$ squares to $-1$, following \cite{shukla, AZ, sven}, we conclude that the current Hamiltonian belongs to DIII symmetry class of Andreev-Altland-Zirnbauer 10-fold classification of random matrix ensembles.

The above discussion applies to all SYK-like models with odd $N$. Now, we specialize to the case of uncolored tensor models. We start by defining an operator $U$ as
\begin{align}
U&=\gamma _2\gamma _4\gamma _6\ldots \gamma _{26}
\end{align} 
From the definition, we can readily see the action of $U$ on gamma matrices as:
\begin{align}
U\gamma _aU^{-1}&=\gamma _a ~~~~~~~~\text{if} ~~ a\in \{2,4,\ldots 26\} \nonumber \\
&=-\gamma _a ~~~~~~\text{if} ~~ a\in \{1,3,\ldots 27,28\}
\end{align} 
As a result, we can show that this operator commutes with $Z$ and $\cal T$ but it anticommutes with $P$. Further, we note that the operator $U$ commutes\footnote{As an aside, we note that the complex conjugation operator $\cal K$ and the operator ${\cal U}_T$ also commute with the Hamiltonian for the reason mentioned in the rest of the paragraph.} with the Hamiltonian \eqref{hamiltonian-repeat}. This is because every term in the Hamiltonian \eqref{hamiltonian-repeat} includes only the gamma matrices from the set $\{\gamma _1, \ldots \gamma _{27}\}$ and has an even number of even\footnote{By even and odd gamma matrices, we mean that the index $a$ in $\gamma _a$ is even and odd respectively.} gamma matrices. This can be proved as follows. Without loss of generality, suppose that the first gamma matrix is an even one and the rest three of them are odd. The assumption of last three gamma matrices being odd implies that $$9a_1+3b_1+c_1+18a_2+6b_2+2c_2$$ is an odd number. This further implies that   the quantity $9a_1+3b_1+c_1$ should be odd which is a contradiction to our initial assumption that the first gamma matrix is an even one. Hence, there exists no terms in the above Hamiltonian which has three odd gamma matrices and one even gamma matrix. A similar argument can be given to prove that there are no terms with one odd gamma matrix and three even gamma matrices. We also note that the operator $U$ is \textit{not} a symmetry of the corresponding SYK Hamiltonian. 

To explain the degeneracies we need to define two more operators. One of them is the fermion number operator $Q$ that is defined to be:
\begin{align}
Q=\sum _{i=1}^{\lceil\frac{N}{2}\rceil}c_i^{\dagger}c_i
\end{align}
where $\lceil x \rceil$ denotes the smallest integer that is greater than or equal to $x$. The \textit{creation} and \textit{annihilation} operators are defined as:
\begin{align}
c_a&=\frac{1}{2}\left(\gamma _{2a-1}+i\gamma _{2a}\right) \\
c_a^{\dagger}&=\frac{1}{2}\left(\gamma _{2a-1}-i\gamma _{2a}\right) 
\end{align}
Although, it is straightforward to see that the Hamiltonian \eqref{hamiltonian-repeat} does not conserve the fermion number $Q$, one can verify that the operator $(-1)^Q$ commutes with the Hamiltonian. Further, we can verify that $Q$ commutes with the operators $P$ and $Z$. But the time reversal operator does not commute with $Q$. Indeed, the action of $\cal T$ on $Q$ is as follows: 
\begin{align}
{\cal T}Q{\cal T}^{-1}&= {\lceil \frac{N}{2} \rceil} -Q
\end{align}  

The last set of conserved operators that we define here are (like $U$) unique to the uncolored tensor models and these operators do not commute with the corresponding SYK Hamiltonian. Before defining these operators explicitly, we notice that in the Hamiltonian \eqref{hamiltonian-repeat}, $a_1$ and $a_2$ fixes\footnote{We note that the similar arguments can be made about $(b_1,b_2)$ and $(c_1,c_2)$.} the range of values that each of the expressions $9a_{1,2}+3b_{1,2}+c_{1,2}$ can take. For instance, choosing $a_1=1$ implies that the first two gamma matrices must belong to the set $\{\gamma _1,\gamma _2\ldots \gamma _9\}$. Extending this example, we can conclude that every term in the Hamiltonian has either zero, two or four gamma matrices belonging to $\{\gamma _1,\gamma _2\ldots \gamma _9\}$ that is obtained by choosing $a_1,a_2\neq 1$ or $a_1=1, a_2\neq 1$; $a_1\neq 1, a_2=1$ or $a_1=a_2=1$ respectively. This statement can be extended to the gamma matrices belonging to the sets  $\{\gamma _{10},\gamma _{11}\ldots \gamma _{18}\}$ and $\{\gamma _{19},\gamma _{20}\ldots \gamma _{27}\}$. 

Thanks to the above observation, the operator $N_1$ defined as
\begin{align}
N_1&=\gamma _1\gamma _2\ldots \gamma _9
\end{align}  
commutes with the Hamiltonian. This can be seen if we note that:
\begin{align}
N_1\gamma _aN_1^{-1}&=\gamma _a ~~~~~~~~\text{if} ~~1\leq a\leq 9 \nonumber \\
&=-\gamma _a ~~~~~~\text{otherwise}
\end{align}
The above observation implies that there will be an even number of negative signs in each of the terms in $N_1HN_1^{-1}$ and hence the operator $N_1$ commutes with the Hamiltonian \eqref{hamiltonian-repeat}. Similarly, we can show that $N_2$ and $N_3$ commute with the Hamiltonian where:
\begin{align}
N_2&=\gamma _{10}\gamma _{11}\ldots \gamma _{18}\\
N_3&=\gamma _{19}\gamma _{20}\ldots \gamma _{27}
\end{align} 
Further, we can readily show the following relations:
\begin{align}
\{N_i,P\}&=0 \\
[N_i,Z]&=0 \\
[N_2,U]&=0 \\
\{N_{1,3},U\}&=0 \\
[N_i,T]&=0 \\
[N_1N_2,Q]&=0
\end{align}

Now that we have defined all the necessary operators, we go on to explain the degeneracies of eigenvalues of the Hamiltonian. To do so, we need to find a set of commuting operators to characterize our eigenstates. We choose the following operators\footnote{We have included `$i$' in the definition of $N'$ to make it Hermitian. } to define the eigenstate basis:
\begin{align}
H, (-1)^Q, P, N'=iN_1N_2 \nonumber
\end{align} 
and write our eigenstates\footnote{We have added an extra label $\alpha$ in the eigenstate to emphasize that there could be other quantum numbers which do not affect the discussion of degeneracy. Note that for this to happen, for a given energy, these quantum numbers must take unique values. An analogous phenomenon happens (for instance) in the case of SYK with $N (\text{mod} ~8) =0$, see \cite{cotler}.} as:
\begin{align}
|E, (-1)^q, p, n',\alpha \rangle \nonumber
\end{align}
Let us start by explaining the degeneracies that are common to uncolored tensor models and the SYK model. Since the operator $Z$ anticommutes with $P$, we get:
\begin{align}
Z~|E, (-1)^q, p, n',\alpha\rangle &=|E, (-1)^q, -p, n',\alpha \rangle
\end{align}
This gives us a two-fold degeneracy. Another two-fold degeneracy \cite{Fu} is obtained because we have ${\cal T}^2=-1$ and that $\cal T$ is an anti-unitary operator. These two features of $\cal T$ imply that for every state $|E, (-1)^q, p, n',\alpha\rangle  $, there exists a state  ${\cal T} |E, (-1)^q, p, n',\alpha\rangle $ that is orthogonal to the original state  (See chapter 2 of \cite{haake} for example).

Now, we describe the four-fold degeneracy that is unique to the uncolored models. The action of the operator $U$ on the eigenstate is given as:
\begin{align}
U|E, (-1)^{q}, p, n',\alpha\rangle &=|E, (-1)^{q+1}, -p, -n',\alpha\rangle
\end{align}
As $U$ squares to unity, another two-fold degeneracy can be understood. The last set of two-fold degeneracies can be understood by noting that 
\begin{align}
N_3~|E, (-1)^q, p, n',\alpha\rangle &=|E, (-1)^{q+1}, -p, n',\alpha\rangle
\end{align}
Since $N_3^2=1$, we have a two-fold degeneracy. Hence, we showed that our model has four independent two-fold degeneracies and thus the uncolored models for $N=27$ has a sixteen fold degeneracy.

We note that the construction of symmetry operators presented here for uncolored models can be extended to Gurau-Witten model. Further, it can be shown that the various symmetry operators we find for the $N=32$ case of Gurau-Witten model do not lead to any degeneracy. This is consistent with the findings in \cite{bala}. 

\section{Random Matrix Ensembles}

In this section, we will generalize the discussion of the spectral mirror symmetry and the time reversal symmetry to the case of arbitrary $n$ for $d=3$ and thus identify \cite{shukla} the symmetry classes to which the Hamiltonian of uncolored model and Gurau-Witten model belong to. We will start our discussion with the uncolored models.

\subsection{Ensembles of the uncolored tensor models}
\label{Uncolored ensembles}

Our goal is to identify the $S$ and ${\cal T}$ operators for these models. To begin with, we assign gamma matrices to fermionic fields (for an arbitrary $n$) as follows:
\begin{align}
\psi ^{ijk}&=\gamma ^{n^2(i-1)+n(j-1)+k}
\end{align}
As argued earlier, we can find a unitary operator $S$ under which the Hamiltonian \eqref{hamiltonian-repeat} of the uncolored tensor models is odd. The unitary operator $S$ acts on the gamma matrices as follows: 
\begin{align}
\label{mirror symmetry}
S\left(\gamma ^{n^2(i-1)+n(j-1)+k}\right)S^{-1}&=\pm \gamma ^{n^2(i-1)+n(k-1)+j}
\end{align}
Operationally, the spectral mirror symmetry operator $S$ exchanges two gamma matrices if $j \neq k$ and does not effect any gamma matrix with $j=k$. Before proceeding to write down the explicit form of the operator $S$, we note that, for $a\neq b$, the operator $\frac{\gamma _a+\gamma _b}{\sqrt{2}}$ exchanges the two gamma matrices $\gamma _a$ and $\gamma _b$ i.e., 
\begin{align}
\frac{1}{2}(\gamma _a+\gamma _b)\gamma _b (\gamma _a+\gamma _b)&=\gamma _a \\
\frac{1}{2}(\gamma _a+\gamma _b)\gamma _a (\gamma _a+\gamma _b)&=\gamma _b
\end{align} 
Once we know an explicit operator that exchanges two gamma matrices, then we expect that the operator $S$ can be constructed by simply taking a product of all such operators. But, to pick the same sign for all gamma matrices under the action of $S$ in \eqref{mirror symmetry}, we also need to include the gamma matrices corresponding to $j=k$ i.e., the operator $S$ can be written\footnote{Note that the ordering of various gammas does not affect our discussions. This is because a change in the ordering of gamma matrices in $S$ may (at most) lead to a change in its sign. This does not affect our results as both the relevant equations \eqref{mirror symmetry} and \eqref{s^2 generic n} are unaffected by a sign change.} as:
\begin{align}
S&=\prod _{j=k} \gamma ^{n^2(i-1)+n(j-1)+k} \prod _{j\neq k}\frac{1}{\sqrt{2}}\left(\gamma ^{n^2(i-1)+n(j-1)+k}+\gamma ^{n^2(i-1)+n(k-1)+j}\right)
\end{align}
Noting that $i,j,k$ take values from $1$ to $n$, we can see that there are $\frac{n^2(n+1)}{2}$ terms in the operator $S$. 

To identify the symmetry class, we need to know the value of $S^2$. Since $S$ is made up of gamma matrices, the value of $S^2$ depends on the number of terms that are present in $S$. For a generic $n$, we can verify that 
\begin{align}
\label{s^2 generic n}
S^2&=(-1)^{\frac{n^2}{8}(n^2-1)(n^2+2n+2)}
\end{align}  

Now, we move on to the time reversal symmetry. Following \cite{xu}, we start by defining an operator $P$ as:
\begin{align}
\label{p operator}
P&=(-i\gamma _1\gamma _2)(-i\gamma _3\gamma _4)\ldots (-i\gamma _{N-1}\gamma _{N}) ~~~~~\text{if}~ N ~\text{is even} \\
&=(-i\gamma _1\gamma _2)(-i\gamma _3\gamma _4)\ldots (-i\gamma _{N}\gamma _{N+1}) ~~~~~\text{if}~ N ~\text{is odd}
\end{align}
where $N$ is the number of independent fermionic fields in the theory. If the number of fermionic fields $N$ is odd, note that we work with the gamma matrices of $SO(N+1)$ instead of that of $SO(N)$. 

The time reversal operator $\cal T$ is defined to be 
\begin{align}
\label{time reversal}
{\cal T}&=P^{(N+2)/2}\gamma _1\gamma _3\ldots \gamma _{N-1}{\cal K} ~~~~~\text{if}~ N ~\text{is even} \nonumber \\
&=P^{(N+3)/2}\gamma _1\gamma _3\ldots \gamma _{N}{\cal K} ~~~~~~~~\text{if}~ N ~\text{is odd}
\end{align}
where $\cal K$ is the complex conjugation operator which acts on the gamma matrices as follows:
\begin{align}
{\cal K}\gamma _a{\cal K}^{-1}&=-(-1)^a\gamma _a
\end{align}
In the present case of uncolored model with $d=3$, we have $N=n^3$. From the explicit form of the time reversal operator \eqref{time reversal}, it is easy to compute ${\cal T}^2$ and the results are summarized in the table \ref{symmetry classes}. The table also includes the symmetry classes of uncolored tensor model Hamiltonians that can be classified\footnote{Note that we can in principle choose the anti-unitary operator $\cal K$ to be our time reversal operator. One major difference between $\cal T$ and $\cal K$ is that the operator $\cal T$ exchanges \cite{xu} the creation and annihilation operators  i.e., 
\begin{align*}
{\cal T}c_i{\cal T}^{-1}&=c^{\dagger}_i \\
{\cal T}c^{\dagger}_i{\cal T}^{-1}&=c_i
\end{align*}
But the complex conjugation operator $\cal K$ does not enjoy this property. Hence we believe that the operator $\cal T$ is more suited to be a time reversal operator as compared to $\cal K$.} using the values of $S^2$ and ${\cal T}^2$. See appendix \ref{AZ classification} for details on the Andreev-Altland-Zirnbauer symmetry  classification based on $S^2$ and ${\cal T}^2$. 

\begin{table}
\centering
\begin{tabular}{c| c c c c c c c c}
$n$ (mod 8) & 0 & 1 & 2 & 3 & 4 & 5 & 6 & 7 \\
 \hline
$S^2 $ & 1 & 1 & -1 & -1 & 1 & -1 & -1 & 1 \\

${\cal T}^2 $ & 1 & 1 & 1 & -1 & 1 & -1 & 1 & 1 \\

Class & BDI & BDI & CI & DIII & BDI & DIII & CI & BDI \\
\end{tabular}
\caption{Symmetry classes of uncolored Hamiltonians for varying $n$ and for $d=3$}
\label{symmetry classes}
\end{table}

\subsection{Symmetry Classes of Gurau-Witten}

A similar analysis can also be performed for the Gurau-Witten model for generic values of $n$ and $d$. In particular, in this section, we wish to identify the symmetry classes of Gurau-Witten Hamiltonian for an arbitrary $d$ and $n$ by computing the values of $S^2$ and ${\cal T}^2$. We start by writing down the Gurau-Witten Hamiltonian \cite{witten} as:
\begin{align}
\label{GW general}
H&=i^{q/2}~\psi _0\psi _1\ldots \psi _d
\end{align}
where $q=d+1$ and each $\psi _a$ has $d$ tensor indices that take values from $1$ to $n$. The gamma matrices corresponding to various fermionic fields are of $SO(qn^d)$ and are given by:
 \begin{align}
 \psi _0^{a_1\ldots a_d}&=\gamma ^{n^{d-1}(a_1-1)+n^{d-2}(a_2-1)+\ldots +a_d} \nonumber \\
 \psi _1^{a_1\ldots a_d}&=\gamma ^{n^{d-1}(a_1-1)+n^{d-2}(a_2-1)+\ldots +a_d+n^d} \nonumber \\
 \vdots \nonumber \\
 \psi _d^{a_1\ldots a_d}&=\gamma ^{n^{d-1}(a_1-1)+n^{d-2}(a_2-1)+\ldots +a_d+dn^d}
 \end{align}
Now that we have written various fermionic fields as gamma matrices, we are ready to explicitly construct the spectral mirror symmetry operator  $S$ and the time reversal operator $\cal T$. 

Let us start with spectral mirror symmetry. Following \cite{bala}, it is easy to see that the product of first $n^d$ gamma matrices anticommute with the Hamiltonian i.e.,
\begin{align}
SHS^{-1}&=-H\\
S&=\gamma _1\gamma _2\ldots \gamma _{n^d}
\end{align}
Note that the operator $S$ is unitary. From the above expression for $S$, we can compute $S^2$ to be:
\begin{align}
S^2&=(-1)^{\frac{n^d}{2}(n^d-1)}
\end{align}

Now, we move on to the time reversal symmetry. Note that the integer $d$ is always odd in the Gurau-Witten model and hence the number of independent fermionic fields $N=qn^d$ is always even and hence the time reversal operator \eqref{time reversal} for the Gurau-Witten model is given by
\begin{align}
{\cal T}&=P^{(N+2)/2}\gamma _1\gamma _3\ldots \gamma _{N-1}{\cal K}
\end{align}
where $N=qn^d$ and $\cal K$ is the complex conjugation operator and the operator $P$ is defined in \eqref{p operator}. First of all, we note that whenever $\frac{q}{2}$ is odd, there is an explicit `$i$' in \eqref{GW general} and thus the Gurau-Witten Hamiltonian is \textit{not}\footnote{A similar argument can also be made regarding the Hamiltonian of SYK model. See \cite{maldacena} for instance.} time-reversal symmetric if $\frac{q}{2}$ is odd. Since the time reversal operator does not exist and since the spectral symmetry operator is unitary, following \cite{shukla, sven}, we conclude that whenever $\frac{q}{2}$ is odd, the Gurau-Witten Hamiltonian belongs to the AIII symmetry class.

For $\frac{q}{2}$ even, we can compute ${\cal T}^2$ as
\begin{align}
{\cal T}^2&=(-1)^{\frac{N}{8}(N-2)}
\end{align}
where $N=qn^d$. Since $\frac{q}{2}$ is even, we write $q=4m$ and thus $N$ is of the form $N=(4m)n^d$. For an even $m$, the above expression implies that ${\cal T}^2=1$. For odd $m$, the value of ${\cal T}^2$ indeed depends on the value of $n^d$. The values of ${\cal T}^2$ and $S^2$ and the symmetry classes of Hamiltonians of Gurau-Witten model for arbitrary $n$ and for even $\frac{q}{2}$ are summarized in tables \ref{symmetry classes GW- odd m} and \ref{symmetry classes GW- even m}.

\begin{table}
\centering
\begin{tabular}{c| c c c c}
$n^d$ (mod 4) & 0 & 1 & 2 & 3  \\
 \hline
$S^2 $ & 1 & 1 & -1 & -1  \\

${\cal T}^2 $ & 1 & -1 & 1 & -1 \\

Class & BDI & CII & CI & DIII  \\
\end{tabular}
\caption{Symmetry classes of Gurau-Witten Hamiltonians with odd $m$ and for arbitrary $n$}
\label{symmetry classes GW- odd m}
\end{table}

\begin{table}
\centering
\begin{tabular}{c| c c c c}
$n^d$ (mod 4) & 0 & 1 & 2 & 3  \\
 \hline
$S^2 $ & 1 & 1 & -1 & -1  \\

${\cal T}^2 $ & 1 & 1 & 1 & 1 \\

Class & BDI & BDI & CI & CI  \\
\end{tabular}
\caption{Symmetry classes of Gurau-Witten Hamiltonians with even $m$ and for arbitrary $n$}
\label{symmetry classes GW- even m}
\end{table}

This completes our discussion of symmetry classes of the random matrix ensembles of the various tensor models. This should be compared to \cite{xu} for SYK. 

\section{Conclusions}

This paper had two goals. 

The first goal was to test the robustness of the salient features that were seen in the spectrum of the colored Gurau-Witten tensor model. We expected to see chaos/random matrix signatures here as well, but we were less sure about the other features, like spectral mirror symmetry, accidental degeneracies, etc.that existed in the spectrum of Gurau-Witten. But we found that the qualitative features of both models are essentially identical. We further explained the degeneracies and spectral mirror symmetry in terms of the symmetries of the Hamiltonian, explicitly.

Our second goal in the paper was to present a classification of the random matrix ensembles that control the behavior of these colored and uncolored tensor models.  A similar analysis for the SYK model (again with $q=4$) was done in \cite{xu} and they found that the random matrix behavior was controlled by the three Wigner-Dyson ensembles, with a Bott periodicity in the size of the fermion. Here on the other we found a more intricate structure controlled by the discrete (pseudo-)symmetries, and that the random matrices belong to the Andreev-Altland-Zirnbauer symmetry classes. The structure is different for the uncolored and colored models, but there is a version of Bott periodicity captured as $n$ changes in both cases.

After analysing $n=3$; $d=3$ case, the next thing one would want to try is to analyse the $n=2$; $d=5$ or $n=4$; $d=3$ case. The former case is not explicitly discussed in \cite{klebanov}, so we will not discuss in detail. We will just mention that in the case of $n=2$; $d=5$, even though the dimensionality of gamma matrices is (barely) tractable with our computing powers, the number of terms in the Hamiltonian is $2^{15}=32768$ and each term is a product of six gamma matrices. Since each of these terms is expected to be sparse in a different way, the final Hamiltonian is likely to be a dense\footnote{ See appendix \ref{sparsity} for details.} matrix.

For the case of $n=4$; $d=3$, the gamma matrices that we need to work with belongs to $SO(64)$. The size of the Hamiltonian is likely too large to implement on standard computers, so we will not address it. The Hamiltonian contains 4096 terms.

For reasons to be elucidated elsewhere, we suspect that the $q\rightarrow \infty$ limit of these models to be the most relevant for quantum gravity. It will also be very interesting to consider the singlet sector of our discussion for holographic purposes: we have ignored this in this paper, but we hope to come back to this question in the future.

\section{Acknowledgments}

CK thanks the organizers and participants of the "Strings Attached" conference at IIT Kanpur where some parts of these results were presented. KVPK thanks the organizers of "Advanced String School" at IOP, Bhubaneswar where part of the work has been completed. 
We acknowledge helpful discussions with Soumyadeep Chaudhuri, Kausik Ghosh, Avinash Raju and P.N. Bala Subramanian.

\appendix

\section{Representation of Gamma matrices}

In this appendix, we construct the representations of Clifford algebra starting with 2-dimensional Pauli matrices ($\sigma _i$). For $m=\left[d/2\right]$ with $d$ being the dimension of the representation, we have (there are $m$ tensor products in each line below):
\begin{align}
\gamma _1&=\sigma _1 \otimes \textbf{1}\otimes \ldots \otimes \textbf{1}  \\
\gamma _2&=\sigma _2 \otimes \textbf{1}\otimes \ldots \otimes \textbf{1}  \\
\gamma _3&=\sigma _3 \otimes \sigma _1 \otimes \textbf{1}\ldots \otimes \textbf{1} \\ 
\gamma _4&=\sigma _3 \otimes \sigma _2 \otimes \textbf{1}\ldots \otimes \textbf{1} \\ 
\vdots \\
\gamma  _{2m-3}&=\sigma _3 \otimes \sigma _3 \ldots \otimes \sigma _3 \otimes \sigma _1 \otimes \textbf{1}\\
\gamma  _{2m-2}&=\sigma _3 \otimes \sigma _3 \ldots \otimes \sigma _3 \otimes \sigma _2 \otimes \textbf{1} \\
\gamma  _{2m-1}&=\sigma _3 \otimes \sigma _3 \ldots \otimes \sigma _3 \otimes \sigma _3 \otimes \sigma _1 \\
\gamma  _{2m}&=\sigma _3 \otimes \sigma _3 \ldots \otimes \sigma _3 \otimes \sigma _3 \otimes \sigma _2 \\
\gamma _{2m+1}&=\sigma _3 \otimes \sigma _3 \ldots \otimes \sigma _3 
\end{align}  
In this paper, we encountered gamma matrices of dimension $d=28$ and thus $m=14$.

\section{Sparsity of a matrix}\label{sparsity}

We define sparsity of a matrix to be
\begin{align}
\text{sparsity}&=\frac{\text{Number of zero elements in the matrix}}{\text{Total number of elements in the matrix}}
\end{align}
The density of a matrix, then, is given by:
\begin{align}
\text{density}&=1-\text{sparsity}
\end{align} 
For the case of $n=3$;$d=3$ uncolored model, the density is $\sim 0.004$ as compared to the density of $\sim 0.0005$ in the $N=32$ Gurau-Witten model considered in \cite{bala} and hence our Hamiltonian is around ten times denser. This higher density can be attributed to the more number of terms\footnote{729 in our case vs 64 in Gurau-Witten.} in our Hamiltonian as compared to \cite{bala}. 

We found ``experimentally" that a useful way to bound the density without explicit knowledge of the Hamiltonian is to take the ratio of the number of terms in the Hamiltonian with its dimensionality (length of row or column).  For the Gurau-Witten case, we found that this is a useful and tight bound, but for the Uncolored Models the bound is too lose. We suspect this is because of the identity piece that shows up in the Hamiltonian in this case. 

\section{Other Symmetries}

Following the construction of spectral mirror symmetry operator $S$, we can construct another operator $S'$ such that
\begin{align}
S'\gamma ^{9i+3j+k-12}S'^{-1}&=\gamma ^{9j+3i+k-12}
\end{align}
The action of operator $S'$ on the Hamiltonian \eqref{hamiltonian-repeat} is given by
\begin{align}
H_{S'}&=S'HS'^{-1} \nonumber \\
&=\sum _{a_1,a_2}\sum _{b_1,b_2}\sum _{c_1,c_2}\psi ^{b_1a_1c_1}\psi ^{b_2a_1c_2}\psi ^{b_1a_2c_2}\psi ^{b_2a_2c_1}
\end{align}
Note that the operator $S'$ is unitary. Rearranging the $\psi ^{ijk}$'s, it is easy to see that $H_{S'}=-H$ i.e., the operator $S'$ anticommutes with the Hamiltonian.  So, the operator $SS'$ is a symmetry of the $n=3$; $d=3$ Hamiltonian. Even though this is a symmetry of the Hamiltonian, we note that the action of $SS'$ on the eigenstates does not lead to any degeneracy. This can be seen if we note that $SS'$ is unitary and \textit{not} Hermitian i.e., it is not an observable. Hence, $SS'$ acting on the eigenstates $|E,(-1)^q,p,n',\alpha\rangle $ just corresponds to rotation\footnote{We thank Avinash Raju for discussions on this point.} of the eigenstates and thus the operator $SS'$ does not lead to a degeneracy.

\section{Andreev-Altland-Zirnbauer Classification}
\label{AZ classification}

In this appendix, we give a brief overview of the Andreev-Altland-Zirnbauer ten-fold classification based on \cite{shukla, sven}. This classification is an extension of the Wigner-Dyson three-fold classification and is based on the unitarity properties of $S$ and also on the values of ${\cal T}^2$ and $S^2$. Here, $\cal T$ is an anti-unitary operator corresponding to time reversal and $S$ is the spectral mirror symmetry operator. 

Spectral mirror symmetry implies that whenever a Hamiltonian has an eigenvalue $E_0+E$, then $E_0-E$ is also a part of the spectrum. Here $E_0$ is the mid-level energy. This implies that the operator $S$ anti commutes with the Hamiltonian i.e.,
\begin{align}
\{H,S\}&=0.
\end{align}

Let us start with the cases where the Hamiltonian has spectral mirror symmetry but does not have time reversal symmetry. In these cases, whenever $S$ is unitary, we can always choose $S^2=1$. But, if $S$ is anti-unitary then we have two different ensembles corresponding to $S^2=+1$ and $S^2=-1$. The three different ensembles that a Hamiltonian that is not invariant under time reversal symmetry can belong to, are:
\begin{itemize}
\item $S$ : Unitary; ~~~~~~~~~$S^2=+1$ ~~~:~ AIII
\item  $S$ : Anti-Unitary; ~~$S^2=+1$ ~~~:~ BD
\item  $S$ : Anti-Unitary; ~~$S^2=-1$ ~~~:~ C
\end{itemize}

Now, we move on to cases where the Hamiltonian is symmetric under both $\cal T$ and $S$. Note that we also need that $\cal T$ and $S$ commute with each other. This leads to four ensembles corresponding to four possibilities among $S^2=\pm 1$ and ${\cal T}^2=\pm 1$ which can be summarized as follows:
\begin{itemize}
\item ${\cal T}^2=+1$;~~$S^2=+1$ ~~~:~ BDI
\item ${\cal T}^2=-1$;~~$S^2=+1$ ~~~:~ CII
\item ${\cal T}^2=+1$;~~$S^2=-1$ ~~~:~ CI
\item ${\cal T}^2=-1$;~~$S^2=-1$ ~~~:~ DIII
\end{itemize}

If the Hamiltonian has no spectral mirror symmetry but is invariant under time reversal, then it belongs to the Wigner-Dyson ensembles depending on the values of ${\cal T}^2$ and are given as:
\begin{itemize}
\item ${\cal T}^2=+1$~~~: AI ~~~(also called Gaussian Orthogonal Ensemble(GOE))
\item ${\cal T}^2=-1$~~~: AII ~~(also called Gaussian Symplectic Ensemble(GSE))
\end{itemize}
When the Hamiltonian does not have either spectral mirror symmetry or time reversal symmetry, it belongs to the symmetry class $A$ (also called Gaussian Unitary Ensemble(GUE)) of the Wigner-Dyson three-fold classification.

\end{document}